# VOLTAGE CONTROLLED AWAKENING OF MEMRISTOR-LIKE DYNAMIC CURRENT-VOLTAGE LOOPS OF FERROELECTRIC TRIGLYCINE SULPHATE

Nicholas V. Morozovsky

*Institute of Physics, National Academy of Sciences of Ukraine,*

*46, Prospekt Nauky, 03028 Kyiv, Ukraine*

E-mail: nicholas.v.morozovsky@gmail.com

**Abstract**

For the study of polarization reversal features in the structure of "mixed ionic-electronic conductor - on - ferroelectric" type, the hydrogen bonded molecular ferroelectric crystal triglycine sulphate (TGS) is used. The dynamic currant-voltage (I-V-) loops of thin TGS plates of polar (010) cut with dissimilar electrodes on the opposite surfaces (vacuum deposited Ag and Ag-paste diluted with a water-ethanol mixture) are investigated. After electroforming in the low-voltage vicinity of coercive voltage, during further cycling, a reversible transformation of I-V-loop shape occurs from typical of non-linear dielectrics to typical of memristive systems and than to typical of ferroelectrics (and vice versa) with a sequential increase (decrease) of drawing voltage amplitude. Such "awakening" and "falling asleep" of I-V-loops of the memristor type is apparently associated with reversible electromigration and accumulation of charged species (e.g., protons), in the hydrated surface layer of TGS as a component of electrochemical capacitor-on-ferroelectric structure. The observed transformation can be explained by considering the coupled ion transfer and polarization reversal taking into account the swinging of the boundary of the built-in chemical inhomogeneity under conditions of linear and/or nonlinear dynamics of ion drift and ionic concentration polarization caused by charged species transfer under the applied voltage.

Memristive two-terminal circuit elements which use ferroelectric (FE) polarization reversal [**1**-**5**] and resistive switching [**6-11**], are the basic ones for non-volatile random access memory (RAM) FERAM and ReRAM devices, respectively. The main difference between FERAM and ReRAM consists in encoding of information by retention of polarization direction for the former (see, e.g., [**4**]) and the resistance value of ion channel for the latter (see, e.g., [**9**]).

In the memristive ReRAM, the switchable conductive atomic bridges [**12**-**15**] and accessors (selector units) [**11**-**16**] are based on the materials with mixed ionic-electronic conduction (MIEC), in which $O^{2-}$ anion and $O^{2+}$ vacancy, as well as $Cu^+$ or $Ag^+$ cation electromigration is used [**8**, **11**, **13**].

For FERAM devices, at least those discussed in the Refs [**1**-**5**], the MIEC-on-FE systems with functionality similar to those used in ReRAM [**11**, **14**, **16**]**,** are not considered**.** At the same time, earlier [**17**, **18**] and recent [**19** - **21**] studies demonstrate the possibility of controlling the orientation of the bulk polarization of ferroelectric film using surface electrochemical processes, including those mediated by water [**19, 21**]. The possible scenarios of polarization reversal for the case of the different species adsorbed from the air with ubiquitous water (O vacancies, O or H adatoms, or OH groups) were considered in the Refs [**17**-**20, 22 - 26**], and the fundamental inseparability of ferroelectric state in the volume and electrochemical state of surface species was shown. Furthermore, in the Ref [**20**] it was pointed out that "Similar phenomena can exist on the interface between ferroelectrics and solid ionic conductors, enabling new classes of ferroelectric devices."

In this regard, direct studies of polarization reversal characteristics on the example of a specific system of the MIEC-on-FE type are of undoubted scientific and applied interest. As far as we know, studies of the characteristics of polarization reversal in MIEC-on-FE systems have not yet been carried out.

For "ad hoc" implementation of such a system, the well-known ferroelectric triglycine sulphate (TGS) can be used.

TGS, $(NH_2CH_2COOH)_3(H_2SO_4)$, is a hydrogen bonded polar molecular crystal [**27 - 29**]. Represented by chemical formula $(NH_3^+CH_2COO^-)(NH_3^+CH_2COOH)_2 \cdot SO_4^{2-}$, TGS may be called glycine-diglycinium sulphate [**27, 31**].

Bulk TGS crystals are suitable for optical second harmonic generation [**30, 31**] and for Fourier transform infrared (IR) instrumentation [**32-34**]. During decades, TGS single crystals, in spite of their hygroscopicity [**32**-**36**], remain preferable for uncooled high-sensitive in wide spectral IR range pyroelectric detectors [**32-39**], which are used for astronomical and earth-based IR thermal imaging [**32, 33, 35, 36**], for Earth surface and atmosphere space IR monitoring [**32, 33, 35, 36**], as well as for Mars [**34**] and Venus [**37**] space missions.

TGS single crystals are grown from an aqueous solution by the method of slow cooling valid even at the space microgravity [**32, 33**] or by the method of slow evaporation of the solvent below or above the Curie temperature ($\approx$ 322 K) [**27, 31**], as well as by vapour diffusion method [**40**]. The reactivity of TGS with water allows etching its surface [**27, 29, 41-43**]. Being water-soluble, TGS is insoluble in ethanol, so the addition of ethanol reduces the speed of TGS solubility etching [**40**].

Earlier it was established [**42**], that after a few seconds of etching the TGS plate in a water bath, regular array of spike-like domains ($\approx$ 30 μm base, $\approx$ 100 μm length) distanced by 100 μm appeared under the etched (010) *b*-surface (see Figs 2.15 and 2.16 in [**27**]). Water-assisted reconstruction of surface topography of TGS crystals has been investigated using humidity-controlled AFM system [**43**]. It was found that the interplay between holes ($\approx$ 0.3 μm) and islands ($\approx$ 3 μm), depending on the relative humidity (RH) in the range (50-70) %, occurs in different ways at two different ends of the domains. The observed interconversion of holes and islands is assumed to be associated with the relocation of glycine molecules in the surface region with the participation of polar water molecules [**43**].

The results of investigations of dielectric [**44, 45**] and ferroelectric [**46**] properties as well as pyroelectric characteristics [**47**] of TGS samples with different electrode materials were interpreted taking into account the effect of ferroelectrically inactive surface layers with low dielectric permittivity, which operate as under-electrode surface capacitors [**45, 47**]. The nature and physical properties of such reconstructed surface layer with reduced effective dielectric permittivity [**29, 38**], often referred to as the "dead layer" [**48**], have been discussed earlier [**29, 38, 48**]. Note that the studies [**44-47**] were performed using the samples with different electrode materials, but with identical electrodes (e.g., Ag-paint in [**44, 45**]). Known studies of current-voltage characteristics under polarization reversal [**49**] and steady-state current-voltage characteristics [**50**] were also carried out for TGS samples with identical electrodes (e.g., Au in [**50**]). To date, as far as we know, the results of similar studies for TGS thin plate samples with dissimilar electrodes have not been published.

For our study, one of the main surfaces of polished $\approx$ 200 μm thin TGS plates of the polar (010) cut was coated with silver electrode deposited in a vacuum. After exposure to ambient conditions (RT, 60-70 % RH, 12 h) the other main surface was coated with a commercial conducting silver paste, pre-dried and diluted with a water-alcohol ($\approx$ 1:10) mixture. Such treatment, due to the solubility of TGS in water, leads to the formation of a specific surface layer at the TGS/Ag-paste interface. This layer, being a hybrid of a surface electrochemical capacitor (ECC) in direct contact with ferroelectric (ECC-on-FE), is similar to the MIEC-on-FE structure mentioned above. ECC, being a more complex system than simple "dead layer", may strongly affect the peculiarities of charge transport. The corresponding information can be obtained by studying current-voltage loops in the range covered coercive voltage.

Dynamic current-voltage (I-V-) loops were registered in the multi-cycle mode. The AC triangle driving voltage amplitude varied in the range of (0 - ±10) V with the 0.5Hz of frequency. The voltage on reference resistor in the sample circuit was displayed using two-channel digital storage oscilloscope (YB54060, Sinometer Instruments). The temporal variations of the I-V-loops during driving voltage cycling were also examined.

The set of current-voltage (I-V) loops obtained under successive increasing and decreasing of the amplitude of driving voltage is presented in the **Figure 1**.

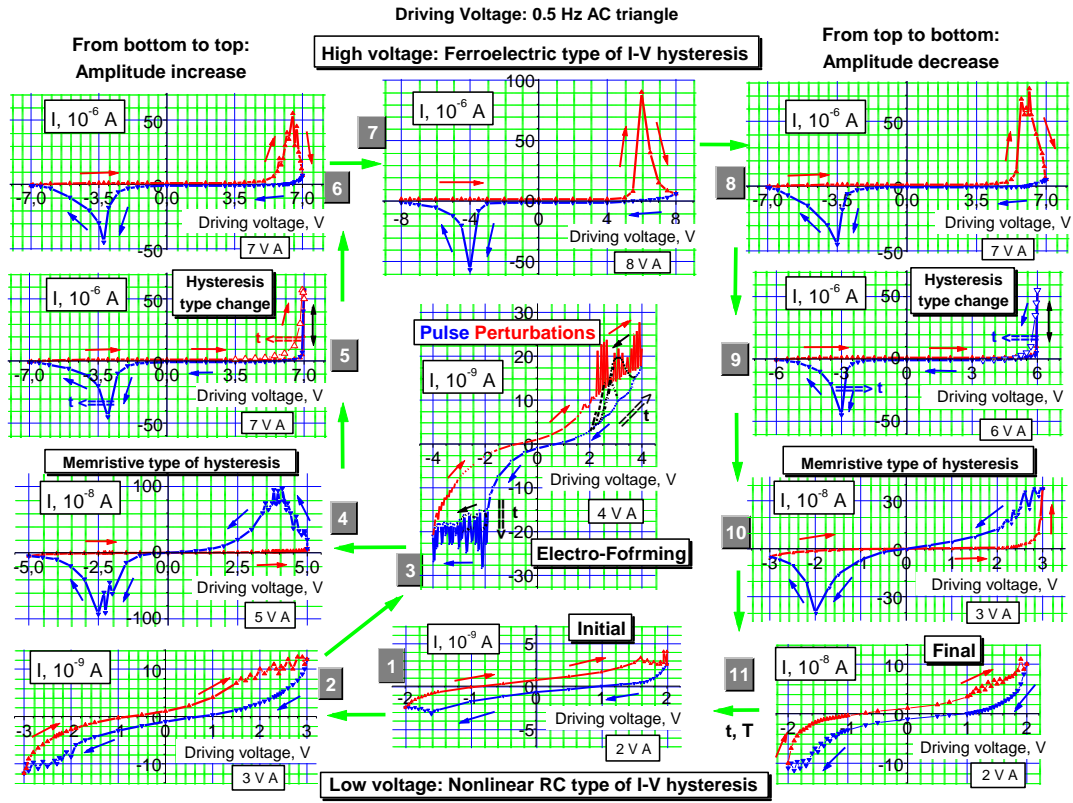

**Figure 1.** The set of current-voltage loops obtained under successive increase (left column, positions 2 - 6, from bottom to top) and decrease (right column, positions 8 - 11, from top to bottom) of the amplitude of driving voltage (0.5 Hz AC triangle; RT).

**1** – Initial current-voltage loop with small Barkhausen jumps;

**2** – Low-voltage uni-hysteretic current-voltage loop characteristic of nonlinear RC circuit;

**3** – Electroforming**:** Variation of type of hysteresis during the "wake-up" stages: Pulse perturbation and appearance of bell-shaped pulses with counter-clockwise cycling direction of hysteresis (see the arrows);

**4** – Near zero crossed double-hysteretic current-voltage loops with memristive type of hysteresis: Opposite cycling direction for positive and negative branches;

**5** – memristor-to-ferroelectric transformation of current-voltage loop: Vanishing of counter-clockwise cycling direction of hysteresis and appearance of clockwise cycling direction of hysteresis;

**6** - **8** – High-voltage pinched current-voltage loops of ferroelectric polarization reversal: Clockwise cycling direction for both positive and negative branches;

**9** – Ferroelectric-to-memristor transformation of current-voltage loop: Vanishing of clockwise cycling direction of hysteresis and appearance of counter-clockwise cycling direction of hysteresis;

**10** – Near zero crossed double-hysteretic current-voltage loops with memristive type of hysteresis: Opposite cycling direction for positive and negative branches;

**11** – Final current-voltage loop characteristic of nonlinear RC circuit.

The shape of initial low-voltage I-V loop (**Fig. 1**, pos. 1) is characteristic of a nonlinear dielectric with losses (nonlinear RC circuit). The small current jumps superimposed on the loop edge are apparently related to Barkhausen effect [**28**, **29**] due to "jerky domain wall motion" [**29**].

Under increase of driving voltage amplitude, the initial I-V loop (**Fig. 1**, pos. 1) is first transformed into uni-hysteretic loop with higher resistive (R) and capacitive (C) nonlinearity and intensity of Barkhausen pulses (**Fig. 1**, pos. 2). Later on, this loop is transformed into double-hysteretic I-V loop with crossing near zero, and with the maxima of the negative and positive branches and with opposite (clockwise and counter-clockwise) cycling direction of each branch (**Fig. 1**, pos. 4).

This transformation is started during the process of "wakening" – electroforming by cycling at constant amplitude of the driving voltage (**Fig. 1**, pos. 3). At the first, "wakening" stage of the electroforming, the intense pulse perturbations of I-V loop arise during the forward run (blue and red arrows in pos. 3 of **Fig**. **1**). At the second, "awakening" stage, these perturbations are followed by bell-shaped pulses with counter-clockwise cycling direction, emerging during the back run (black curves and black double arrows in pos. 3 of **Fig**. **1**). At the third, "waking" stage, these pulses are transformed into the stable maximum with the counter-clockwise cycling direction (blue arrows in pos. 4 of **Fig**. **1**).

Under subsequent increase of driving voltage amplitude, the hysteresis of the branch with counter-clockwise cycling direction vanishes (see black two-edged arrow in pos. 5 of **Fig. 1**), and then turns into a hysteresis with clockwise cycling direction (see red double arrow in pos. 5 of **Fig. 1**,). As a result, the I-V-loop is transformed into an almost symmetric and pinched loop of ferroelectric polarization reversal with opposite maxima and clockwise cycling direction of both branches (**Fig. 1**, pos. 6 and 7). This transformation is accompanied by decrease of the intensity of Barkhausen pulses (cf. pos. 4-7 in **Fig. 1**) and by some increase of the coercive field. The last is characteristic of TGS [**29**].

During subsequent decrease of driving voltage amplitude, the shape of the I-V loop is changed in the reverse order (**Fig. 1**, pos. 8-11). The back reversion of cycling direction, the "falling asleep", starts at lower driving voltage amplitude than the forward one (cf. pos. 5 and 9 in **Fig. 1**). For the final loop (**Fig. 1**, pos. 11), the degree of RC nonlinearity is higher than for initial one (**Fig. 1**, pos. 1) and is restored at short time heating (RT + 20 ºC).

The general shape of "awakened" zero-crossed I-V-loops (**Fig. 1**, pos. 4 and 10) is close to the memristive one with bipolar type of switching [**8**]. In this case, zero-crossed I-V loop has asymmetric negative and positive branches with opposite maxima and opposite its cycling direction (clockwise and counter-clockwise) (see blue and red arrows in pos. 4 and 10 of **Fig. 1**). Such type of zero-crossed I-V-loops is characteristic of memristive tunnel junctions based on ferroelectric materials $BaTiO_3$ [**1**, **2**] and $Pb(Zr_{0.2}Ti_{0.8})O_3$ [**3**], to Pt/organic monolayer/Ti devices [**51**], as well as to electrochemical systems with MIEC [**52**] and to $Au-Ti/TiO_2/Ti-Au$ system with nonlinear electromigration of oxygen vacancies [**53**].

Apparently, the performed treatment transforms the near-surface TGS layer beneath the Ag-paste electrode into a proton rich ion-conducting quasi-2D electrochemical cell. Polar water molecules H-O-H (dipole moment 1.84 D) distort the TGS hydrogen bond network and cause changes in the position of polar glycine molecules. (The dipole moment of glycine ($NH_2CH_2COOH$) estimated for an aqueous 1 M solution is 15.7D [**54**] and calculated for a glycine zwitterion ($NH_3^+CH_2COO^-$) is 11.9D [**55**]). The rearrangement of glycine molecules, possibly similar to proposed for the humidity controlled β-to-γ phase transformation in glycine microcrystals [**56**] or to the neutral-to-ionic transition in other molecular crystals [**57**], gives rise to ionic charge-transfer in the near-surface TGS layer. In particular, the proton transfer between $NH_3^+CH_2COOH$ - $NH_3^+CH_2COO^-$ and $NH_3^+CH_2COOH$ - $SO_4^{2-}$ layers of TGS [**27**, **42**] should realise with participation of longest (weakest) N-H-O bods triggered by H-O-H water molecules.

Due to reversible electromigration and accumulation of charged species (including eventual $Ag^+$ transfer), such an ECC-on-FE structure demonstrates a reversible "awakening" and "falling asleep" of dynamic memristor-like I-V loops: the transformation from characteristic of non-linear dielectric into memristor-like and than into ferroelectric (and vice versa) with a sequential increase (decrease) of drawing voltage amplitude.

The theoretical model of the memristive I-V-loops was proposed for the case of swinging of the boundary of built-in chemical inhomogeneity under conditions of linear dynamics of ion drift [**6**, **7**, **16**] and was recently developed for the case of nonlinear dynamics of ion drift [**53**].

An explanation of the observed transformation can be given when considering the coupled ion transfer and polarization reversal [**58**, **59**] taking into account the findings of [**6**, **7**, **16**] and [**53**], as well as the effects of ion concentration polarization [**16**, **60**] caused by protons $H^+$ transfer under external voltage.

Obtained results stimulate further researches and suggest the unforeseen possibilities for applications of hydrogen bounded molecular ferroelectrics in the logic, memory and neuromorphic systems. Also, the results obtained shed light on how the "dead" layer in thin ferroelectric plates can be converted into effective one, at least for ferroelectrics of TGS family.

**References**

1. A. Chanthbouala, V. Garcia, R. O. Cherifi, K. Bouzehouane, S. Fusil, X. Moya, S. Xavier, H. Yamada, C. Deranlot, N. D. Mathur, M. Bibes, A. Barthélémy, J. Grollier, "A ferroelectric memristor". Nature Materials, *11*(10), 860–4 (**2012**); DOI: 10.1038/NMAT3415.
2. D. J. Kim, H. Lu, S. Ryu, C.-W. Bark, C.-B. Eom, E. Y. Tsymbal, and A. Gruverman, "Ferroelectric tunnel memristor". Nano Lett., *12*, 5697-702 (**2012**); dx.doi.org/10.1021/nl302912t.
3. A. Quindeau, D. Hesse, and M. Alexe, "Programmable ferroelectric tunnel memristor". Frontiers in Phys. Condens. Mater. Phys., *2*, Art. 7, 1-5 (**2014**); doi: 10.3389/fphy.2014.00007.
4. V. Garcia & M. Bibes, "Ferroelectric tunnel junctions for information storage and processing". Nature Communications, *5*, 4289 (12 pp.) (**2014**); DOI: 10.1038/ncomms5289.
5. Z. Wang, W. Zhao, W. Kang, Y. Zhang, J.-O. Klein, D. Ravelosona, and C. Chappert, "Compact modelling of ferroelectric tunnel memristor and its use for neuromorphic simulation". Appl. Phys. Lett., *104*(5), 053505-1-5 (**2014**); http://dx.doi.org/10.1063/1.4864270.
6. D. B. Strukov, G. S. Snider, D. R. Stewart and R. S. Williams "The missing memristor found". Nature Letters, *453*, 80–3 (**2008**); doi:10.1038/nature06932.
7. D. B. Strukov, J. L. Borghetti, and R. S. Williams, "Coupled ionic and electronic transport model of thin-film semiconductor memristive behavior". Small, *5*(9), 1058-63 (**2009**); DOI: 10.1002/smll.200801323.
8. R. Waser and M. Aono, "Nanoionics-based resistive switching memories", Nature Materials, *6*, 833-40, **2007**.
9. K. M. Kim, D. S. Jeong and Ch. S. Hwang, "Nanofilamentary resistive switching in binary oxide system; a review on the present status and outlook". Nanotechnology, *22*, 254002 (17pp) (**2011**); doi:10.1088/0957-4484/22/25/254002.
10. P. Mazumder, Sung Mo Kang, R. Waser, "Memristors: Devices, Models, and Applications". Proc. IEEE, *100*(6), 1911-9 (**2012**); DOI: 10.1109/JPROC.2012.2190812.
11. G. W. Burr, R. S. Shenoy, K. Virwani, P. Narayanan, A. Padilla, B. Kurdi, and H. Hwang, "Access devices 3D crosspoint memory". J. Vac. Sci. Technol. B, *32*(4), 040802-1-23 (**2014**); http://dx.doi.org/10.1116/1.4889999.
12. I. Valov, R. Waser, J. R. Jameson and M. N. Kozicki. "Electrochemical metallization memories: fundamentals, applications, prospects". Nanotechnology, *22*(25), 1-22 254003 (22 pp.) (**2011**); DOI: 10.1088/0957-4484/22/25/254003.
13. A. Wedig, M. Luebben, D.-Y. Cho, M. Moors, K. Skaja, V. Rana, T. Hasegawa, K. K. Adepalli, B. Yildiz, R. Waser, & I. Valov: "Nanoscale cation motion in $TaO_x$, $HfO_x$ and $TiO_x$ memristive systems". Nat. Nano., *11*, 67-74 (**2016**); doi: 10.1038/nnano.2015.221.
14. D. Ielmini and R. Waser (Eds). "Resistive Switching: From Fundamentals of Nanoionic Redox Processes to Memristive Device Applications", 1-st Edition. Wiley-VCH Verlag GmbH & Co. KGaA (**2016**).
15. M. Aono (Ed.), "Atomic Switch: From Invention to Practical Use and Future Prospects". Springer Nature, 266 pp. (**2020**); https://doi.org/10.1007/978-3-030-34875-5.
16. P. Meuffels and R. Soni. "Fundamental issues and problems in the realization of memristors", Forschungzentrum Jlich GmbH, Peter Grnberg Institut, Germany and Nanoelektronik, Technische Fakultt, Christian- Albrechts-Universitt zu Kiel, Germany (**2012**); arXiv:1207.7319v1 [cond-mat.mes-hall].
17. R. V. Wang, D. D. Fong, F. Jiang, M. J. Highland, P. H. Fuoss, Carol Thompson, A. M. Kolpak, J. A. Eastman, S. K. Streiffer, A. M. Rappe, and G.B. Stephenson, "Reversible chemical switching of a ferroelectric film". Phys. Rev. Lett., *102*, 047601-1-4 (**2009**).
18. N. C. Bristowe, M. Stengel P. B. Littlewood, J. M. Pruneda, and E. Artacho, "Electrochemical ferroelectric switching: Origin of polarization reversal in ultrathin films". Phys. Rev. B, *85*(2), 024106-1-7 (**2012**); DOI: 10.1103/PhysRevB.85.024106.
19. C. Blaser, & P. Paruch, "Subcritical switching dynamics and humidity effects in nanoscale studies of domain growth in ferroelectric thin films". New Journal of Physics, *17*(1), 013002-1-8 (**2015**); doi:10.1088/1367-2630/17/1/013002.
20. Sang Mo Yang, A. N. Morozovska, R. Kumar, E. A. Eliseev, Ye Cao, L. Mazet, N. Balke, S. Jesse, Rama K. Vasudevan, C. Dubourdieu and S. V. Kalinin, "Mixed electrochemical–ferroelectric states in nanoscale ferroelectrics". Nature Physics, *13*(8), 812–18 (**2017**); doi:10.1038/nphys4103.
21. N. Domingo, I. Gaponenko, K. Cordero-Edwards, N. Stucki, V. Pérez-Dieste, C. Escudero, E. Pach, A. Verdaguer, and P. Paruch, "Surface charged species and electrochemistry of ferroelectric thin films". Nanoscale, *11*, 17920-30 (**2019**); DOI: 10.1039/C9NR05526F.

22. G. B. Stephenson and M. J. Highland, "Equilibrium and stability of polarization in ultrathin ferroelectric films with ionic surface compensation". Phys. Rev. B, *84*, 064107 (**2011**).
23. M. J. Highland, T. T. Fister, D. D. Fong, P. H. Fuoss, C. Thompson, J. A. Eastman, S. K. Streiffer, and G. B. Stephenson, "Equilibrium polarization of ultrathin $PbTiO_3$ with surface compensation controlled by oxygen partial pressure". Phys. Rev. Lett., *107*(18), 187602-1-5 (**2011**).
24. A. N. Morozovska, E. A. Eliseev, N. V. Morozovsky, and S. V. Kalinin, "Ferroionic states in ferroelectric thin films". Phys. Rev. B, *95*, 195413-1-17 (2017); DOI: 10.1103/PhysRevB.95.195413.
25. A. N. Morozovska, E. A. Eliseev, A. I. Kurchak, N. V. Morozovsky, R. K. Vasudevan, M. V. Strikha, and S. V. Kalinin, "Effect of surface ionic screening on the polarization reversal scenario in ferroelectric thin films: Crossover from ferroionic to antiferroionic states". Phys. Rev. B, *96*, 245405-1-14 (**2017**).
26. A. N. Morozovska, E. A. Eliseev, I. S. Vorotiahin, M. V. Silibin, S. V. Kalinin, and N. V. Morozovsky, "Control of polarization reversal temperature behavior by surface screening in thin ferroelectric films". Acta Materialia, *160*, 57-71 (**2018**); https://doi.org/10.1016/j.actamat.2018.08.041.
27. F. Jona, G. Shirane, "Ferroelectric Crystals", Pergamon Press: London (**1962**).
28. M. E. Lines, A. M. Glass, "Principles and Applications of Ferroelectrics and Related Materials", Clarendon Press, Oxford (**1977**); University Press, Oxford (**2001**).
29. A. K. Tagantsev, L. E. Cross, J. Fousek, "Domains in Ferroic Crystals and Thin Films", Springer, New York - Dordrecht - Heidelberg - London (**2010**).
30. V. S. Suvorov, A. S. Sonin, "Second harmonic generation in triglycine sulphate crystals". Zh. Eksp. Teor. Fiz., *54*(4), 1044-50 (**1967**) (Soviet Physics JETP, *27*(4), 1968).
31. P. R. Deepti, J. Shanti, "Structural and optical studies of potential ferroelectric crystal: KDP doped TGS". J. Sci. Res., *6*(1), 1-9 (**2014**). doi: http://dx.doi.org/10.3329/jsr.v6i1.16584.
32. R. B. Lal, A. K. Batra, "Growth and properties of triglycine sulfate(TGS) crystals: Review". Ferroelectrics, *142*(1-4), 51-82 (**1993**); DOI: 10.1080/00150199308237884.
33. R. B. Lal, A. K. Batra, "Triglycine sulfate (TGS) crystals for pyroelectric infrared detecting devices". SPIE, Infrared Technology XX, 2269, 380-5 (**1994**).
34. P. R. Christensen, G. L. Mehall, S. H. Silverman, S. Anwar, G. Cannon, N. Gorelick, R. Kheen, T. Tourville, D. Bates, S. Ferry, T. Fortuna, J. Jeffryes, W. O'Donnell, R. Peralta, T. Wolverton, D. Blaney, R. Denise, J. Rademacher, R. V. Morris, and S. Squyres, "Miniature Thermal Emission Spectrometer for the Mars Exploration Rovers". Journal of Geophysical Research, *108*(E12), 8064 (23 pp.), (**2003**); doi:10.1029/2003JE002117.
35. S. B. Lang, "Pyroelectricity: From Ancient Curiosity to Modern Imaging Tool". Physics Today, *58*(8), 31-6 (**2005**).
36. M. D. Aggarwal, A. K. Batra, P. Guggilla, M. E. Edwards, J. R. Currie, Jr., "Pyroelectric materials for uncooled infrared detectors: processing, properties, and applications". NASA/TM—2010–216373, George C. Marshall Space Flight Center, **2010**; https://www2.sti.nasa.gov.
37. F. M. Taylor, F. E. Vescelus, J. R. Locke, G. T. Foster, P. B. Forney, R. Beer, J. T. Houghton, F. T. Delderfield, "Infrared radiometer for the Pioneer-Venus orbiter". Appl. Opt., *18*(23), 3893-900 (**1979**).
38. J. C. Burfoot, G. W. Taylor, "Polar Dielectrics and Their Applications", Macmillan Press: London (**1979**).
39. S. B. Lang, D. K. Das-Gupta, "Pyroelectricity: Fundamentals and applications", in "Handbook of Advanced Electronic and Photonic Materials and Devices". Academic Press, San Diego, 1–55 (2001).
40. J. M. Hudspeth, D. J. Goossens, "Vapour diffusion growth and characterisation of fully deuterated triglycine sulphate $(ND_2CD_2COOD)_3D_2SO_4$". Journal of Crystal Growth, *338*(1), 177-180 (**2012**); http://dx.doi.org/10.1016/j.jcrysgro.2011.11.002.
41. V. P. Konstantinova, I. M. Sil'vestrova, V. A Yurin, "Twinning and dielectric behavior of crystals of triglycine sulfate". Kristallografiya, *4*(1), 125-9 (**1959**).
42. A. G. Chynoweth, W. L. Feldman, "Ferroelectric domain delineation in triglycine sulphate and domain arrays produced by thermal shocks". J. Phys. Chem. Solids., *15*(3-4), 225-33 (**1960**); https://doi.org/10.1016/0022-3697(60)90246-8.
43. S. Balakumar, H. C. Zeng, "Water-assisted reconstruction on ferroelectric domain ends of triglycine sulfate $(NH_2CH_2COOH)_3·H_2SO_4$ crystals". J. Mater. Chem., *10*(3), 651-6 (**2000**); DOI:10.1039/A907937H.
44. A. Mansingh, S. S. Bava, "The effect of surface layer on the dielectric behaviour of triglycine sulphate". J. Phys. D: Appl. Phys., *7*(15), 2097-2104 (**1974**); DOI https://doi.org/10.1088/0022-3727/7/15/314.
45. M. S. Tsedrik, G.A. Zaborovski, N. V. Ulasen', "Surface layer effect on variation of ε and tanδ of triglycine-sulfate crystals". Reports of Academy of Science of Byelorussia, *19*(6), pp. 498-501 (**1975**).

46. A. Mansingh, S. S. Bava, "The effect of surface layer on the spontaneous polarization of triglycine sulphate". J. Phys. D: Appl. Phys., *8*(9), 1156-61 (**1975**); DOI: https://doi.org/10.1088/0022-3727/8/9/022.
47. Z. Málek, J. Janta, G. Chanussot. "The influence of surface effects on the pyroelectric behaviour of T.G.S. close to the phase transition". Journal de Physique Colloques, *33*(C2), C2-233-4 (**1972**); http://dx.doi.org/10.1051/jphyscol:1972280.
48. Li-Wu Chang, M. Alexe, J. F. Scott, J. M. Gregg, "Settling the ''Dead Layer'' Debate in Nanoscale Capacitors". Adv. Mater., *21*, 4911-4 (**2009**); DOI: 10.1002/adma.200901756.
49. M. P. Michailov, S. R. Stoyanov, E. M. Michailova, "Influence of the bias electric field on the current-voltage characteristics of triglycine sulphate doped with L-α-alanine". Universite de Plovdiv "Paissi Hilendarski". Travaux Scientifiques, Physique, *19*(2), 171-9 (**1981**).
50. W. Osak, "Charge transport and relaxation in triglycine sulphate (TGS)". Z. Naturforsch, *52a*, 621-8 (**1997**).
51. D. R. Stewart, D. A. A. Ohlberg, P. A. Beck, Y. Chen, and R. Stanley Williams, J. O. Jeppesen, K. A. Nielsen, J. Fraser Stoddart, "Molecule-independent electrical switching in Pt/organic monolayer/Ti devices". Nano Lett. *4*, pp. 133–6 (**2004**); https://doi.org/10.1021/nl034795u.
52. K. MacVittie, E. Katz, "Electrochemical systems with memimpedance properties", The Journal of Physical Chemistry C, *117*(47), 24943-7 (**2013**); DOI: 10.1021/jp409257v.
53. S. Alialy, K. Esteki, M. S. Ferreira, J. J. Boland, C. Gomes da Rocha, "Nonlinear ion drift-diffusion memristance description of $TiO_2$ RRAM devices". Nanoscale Advances, *2*, 2514-24 (**2020**); doi:10.1039/d0na00195c.
54. M. W. Aaron, E. H Grant, "Dielectric and viscosity studies on the dipeptides of alanine and glycine". Brit. J. Appl. Phys., *18*(7), 957-63 (**1967**). DOI: 10.1088/0508-3443/18/7/311.
55. T. Sato, R. Buchner, Š. Fernandez, A. Chiba, W. Kunz, "Dielectric relaxation spectroscopy of aqueous amino acid solutions: dynamics and interactions in aqueous glycine". Journal of Molecular Liquids, *117*(1-3), 93–8 (**2005**), doi:10.1016/j.molliq.2004.08.001.
56. D. Isakov, D. Petukhova, S. Vasilev, A. Nuraeva, T. Khazamov, E. Seyedhosseini, P. Zelenovskiy, V. Ya. Shur, and A. L. Kholkin, "In situ observation of the humidity controlled polymorphic phase transformation in glycine microcrystals". Cryst. Growth Des., *14*, 4138−42 (**2014**); dx.doi.org/10.1021/cg500747x.
57. M. Buron-Le Cointe, E. Collet, B. Toudic, P. Czarnecki and H. Cailleau, "Back to the structural and dynamical properties of neutral-ionic phase transitions". Crystals, *7*, 285 (37 pp.) (**2017**); doi:10.3390/cryst7100285.
58. A. N. Morozovska, E. A. Eliseev, O. V. Varenyk, Y. Kim, E. Strelcov, A. Tselev, N. V. Morozovsky, S. V. Kalinin, "Nonlinear space charge dynamics in mixed ionic-electronic conductors: Resistive switching and ferroelectric-like hysteresis of electromechanical response". J. Appl. Phys., *116*, 066808-1-11 (**2014**); doi:10.1063/1.4891346. (See **Fig. 8b**)
59. A. N. Morozovska, E. A. Eliseev, P. S. Sankara Rama Krishnan, A. Tselev, E. Strelcov, A. Borisevich, O. V. Varenyk, N. V. Morozovsky, P. Munroe, S. V. Kalinin, V. Nagarajan, "Defect thermodynamics and kinetics in thin strained ferroelectric films: The interplay of possible mechanisms". Phys. Rev. B, *89*, 054102-1-10 (**2014**); DOI: 10.1103/PhysRevB.89.054102. (See **Fig. 7d**)
60. K. Kontturi, J. Manzanares, L. Murtomäki, "Effect of concentration polarization on the current-voltage characteristics of ion transfer across ITIES". Elecrrochimica Acta, *40*(18), 2979-84 (**1995**); doi.org/10.1016/0013-4686(95)00231-3.